\documentclass[epj]{svjour}
\usepackage{graphicx}

\def\mb#1{\setbox0=\hbox{$#1$}\kern-.025em\copy0\kern-\wd0
\kern-0.05em\copy0\kern-\wd0\kern-.025em\raise.0233em\box0}

\begin{document}
   \title{Jeans type instability for a chemotactic model of cellular aggregation}

 \author{P.H. Chavanis}

\institute{ Laboratoire de Physique Th\'eorique, Universit\'e Paul
Sabatier, 118 route de Narbonne 31062 Toulouse, France\\
\email{chavanis@irsamc.ups-tlse.fr}}

\titlerunning{}

   \date{To be included later }

   \abstract{We consider an inertial model of chemotactic aggregation
   generalizing the Keller-Segel model and we study the linear
   dynamical stability of an infinite and homogeneous distribution of
   cells (bacteria, amoebae, endothelial cells,...) when inertial
   effects are accounted for. These inertial terms model cells
   directional persistance. We determine the condition of instability
   and the growth rate of the perturbation as a function of the cell
   density and the wavelength of the perturbation. We discuss the
   differences between overdamped (Keller-Segel) and inertial
   models. Finally, we show the analogy between the instability
   criterion for biological populations and the Jeans instability
   criterion in astrophysics.   }

   \maketitle
%

\section{Introduction}
\label{sec_introduction}

The self-organization of biological cells (bacteria, amoebae,
endothelial cells,...) or even insects (like ants) due to the
long-range attraction of a chemical (pheromone, smell, food,...) 
produced by the organisms themselves is a long-standing problem in
physical sciences \cite{murray}. This process is called
chemotaxis. The chemotactic aggregation of biological populations is
usually studied in terms of the Keller-Segel model
\cite{ks}:
\begin{eqnarray}
\label{intro1} \xi\frac{\partial\rho}{\partial t}=\nabla \cdot
(D_{2}\nabla\rho-D_{1}\nabla c),
\end{eqnarray}
\begin{eqnarray}
\label{intro2} \epsilon\frac{\partial c}{\partial t}=-k(c)c+\rho
f(c)+D\Delta c,
\end{eqnarray}
which consists in a drift-diffusion equation (\ref{intro1}) governing
the evolution of the density of cells $\rho({\bf r},t)$ coupled to a
diffusion equation (\ref{intro2}) involving terms of source and
degradation for the secreted chemical $c({\bf r},t)$. The chemical is
produced by the organisms (cells) at a rate $f(c)$ and is degraded at
a rate $k(c)$. It also diffuses according to Fick's law with a
diffusion coefficient $D$. The concentration of cells changes as a
result of an oriented chemotactic motion in a direction of a positive
gradient of the chemical and a random motion analogous to
diffusion. In Eq. (\ref{intro1}), $D_2(\rho,c)$ is the diffusion
coefficient of the cells and $D_1(\rho,c)$ is a measure of the
strength of the influence of the chemical gradient on the flow of
cells. These coefficients depend a priori on the concentration of
cells and on the concentration of the chemical. The Keller-Segel model
is able to reproduce the chemotactic aggregation (collapse) of
biological populations when the attractive drift term $D_{1}\nabla c$
overcomes the diffusive term $D_{2}\nabla\rho$. 

However, recent experiments of {\it in vitro} formation of blood
vessels show that cells randomly spread on a gel matrix autonomously
organize to form a connected vascular network that is interpreted as
the beginning of a vasculature \cite{gamba}. This phenomenon is
responsible of angiogenesis, a major actor for the growth of
tumors. These networks cannot be explained by the {\it parabolic
model} (\ref{intro1})-(\ref{intro2}) that leads to pointwise
blow-up. However, they can be recovered by {\it hyperbolic models}
that lead to the formation of networks patterns that are in good
agreement with experimental results. These models take into account
inertial effects and they have the form of hydrodynamic equations
\cite{gamba}:
\begin{equation}
\label{ga1} {\partial\rho\over\partial t}+\nabla\cdot (\rho{\bf u})=0,
\end{equation}
\begin{equation}
\label{ga2} \frac{\partial {\bf u}}{\partial t}+({\bf u}\cdot
\nabla){\bf u}=-{1\over\rho}\nabla p+\nabla c,
\end{equation}
\begin{equation}
\label{ga3}{\partial c\over\partial t}=-k c+f\rho+D_{c}\Delta c.
\end{equation}
The inertial term models cells directional persistance and the general
density dependent pressure term $-\nabla p(\rho)$ can take into
account the fact that the cells do not interpenetrate. In these
models, the particles concentrate on lines or filaments
\cite{gamba,filbet}. These structures share some analogies with the
formation of ants' networks (due to the attraction of a pheromonal
substance) and with the large-scale structures in the universe that
are described by similar hydrodynamic (hyperbolic) equations, the
Euler-Poisson system \cite{peebles}. The similarities between the
networks observed in astrophysics (see Figs 10-11 of
\cite{vergassola}) and biology (see Figs 1-2 of \cite{gamba}) are
striking.

In order to make the connection between the parabolic model
(\ref{intro1})-(\ref{intro2}) and the hyperbolic model
(\ref{ga1})-(\ref{ga3}), we consider a model of the form
\begin{eqnarray}
\label{intro3} \frac{\partial\rho}{\partial t}+\nabla\cdot (\rho {\bf
u})=0,
\end{eqnarray}
\begin{eqnarray}
\label{intro4} \rho\left \lbrack \frac{\partial {\bf u}}{\partial
t}+({\bf u}\cdot \nabla){\bf
u}\right\rbrack=-D_{2}\nabla\rho+D_{1}\nabla c-\xi\rho {\bf u},
\end{eqnarray}
\begin{eqnarray}
\label{intro5} \epsilon\frac{\partial c}{\partial t}=-k(c)c+\rho
f(c)+D\Delta c,
\end{eqnarray}
including a friction force $-\xi\rho {\bf u}$. This type of damped
hydrodynamic equations was introduced in \cite{csr,gt} at a general level.
This inertial model takes into account the fact that the particles do
not respond immediately to the chemotactic drift but that they have
the tendency to continue in a given direction on their own. However,
after a relaxation time of order $\xi^{-1}$, their velocity will be
aligned with the chemotactic gradient. This is modeled by an effective
friction force in Eq. (\ref{intro4}) where the friction coefficient
$\xi\sim \tau^{-1}$ is interpreted as the inverse of the relaxation
time. This term can also represent a physical friction of the
organisms against a fixed matrigel. In the strong friction limit
$\xi\rightarrow +\infty$, or for large times $t\gg\xi^{-1}$, one can
formally neglect the inertial term in Eq. (\ref{intro4}) and obtain
\cite{gt,crrs,virial}:
\begin{eqnarray}
\label{intro6} \rho {\bf u}=-\frac{1}{\xi}\left
(D_{2}\nabla\rho-D_{1}\nabla c\right )+O(\xi^{-2}).
\end{eqnarray}
Substituting this relation in Eq. (\ref{intro3}), we recover the
parabolic Keller-Segel model (\ref{intro1})-(\ref{intro2}). Therefore,
the Keller-Segel model can be viewed as an overdamped limit of a
hydrodynamic model involving a friction force. Alternatively,
neglecting the friction force $\xi=0$, we recover the hydrodynamic
model introduced in \cite{gamba}.

In this paper, we study the linear dynamical stability of an infinite
and homogeneous distribution of cells with respect to the inertial
model (\ref{intro3})-(\ref{intro5}). We determine the condition of
instability and the growth rate of the perturbation, and discuss the
differences with the results obtained with the Keller-Segel model
(\ref{intro1})-(\ref{intro2}). We also discuss some analogies with the
dynamical stability of self-gravitating systems. Indeed, there are
many analogies between the chemotactic aggregation of biological
populations and the dynamics of self-gravitating Brownian particles
\cite{crrs}. In particular, the Keller-Segel model
(\ref{intro1})-(\ref{intro2}) is similar to the Smoluchowski-Poisson
system \cite{crs} and the hydrodynamic equations
(\ref{intro3})-(\ref{intro5}) are similar to the damped Euler
equations of Brownian particles \cite{gt,virial}.  The main difference
between biological systems and self-gravitating Brownian particles is
that the Poisson equation in the gravitational problem is replaced by
a more general field equation (\ref{intro5}) taking into account the
specificities of the biological problem. Owing to this analogy, we
shall discuss the relation between the instability criterion of
biological populations and the Jeans instability criterion
\cite{jeans} in astrophysics.

\section{Instability criterion for biological populations}
\label{sec_instinert}

\subsection{The dispersion relation}
\label{sec_dis}

We consider an infinite and homogeneous stationary solution of Eqs.
 (\ref{intro3})-(\ref{intro5}) with ${\bf u}={\bf 0}$, $\rho={\rm Cst.}$ and $c={\rm Cst.}$
such that
\begin{eqnarray}
\label{dis1}
k(c)c=f(c) \rho.
\end{eqnarray}
Linearizing the equations around this stationary solution, we get
\begin{eqnarray}
\label{dis2}
\frac{\partial \delta\rho}{\partial t}+\rho \nabla\cdot \delta {\bf u}=0,
\end{eqnarray}
\begin{eqnarray}
\label{dis3}
\rho \frac{\partial \delta{\bf u}}{\partial t}=-D_{2}\nabla\delta\rho+D_{1}\nabla \delta c-\xi\rho \delta{\bf u},
\end{eqnarray}
\begin{eqnarray}
\label{dis4}
\epsilon\frac{\partial \delta c}{\partial t}=(f'(c)\rho-\overline{k})\delta c+ f(c)\delta\rho+D\Delta \delta c,
\end{eqnarray}
where we have set $\overline{k}=k(c)+c k'(c)$. Eliminating the velocity between Eqs. (\ref{dis2}) and (\ref{dis3}), we obtain
\begin{eqnarray}
\label{dis5}
\frac{\partial^{2}\delta\rho}{\partial t^{2}}+\xi\frac{\partial\delta\rho}{\partial t}=D_{2}\Delta\delta\rho-D_{1}\Delta\delta c.
\end{eqnarray}
Looking for solutions of the form $\delta\rho\sim \delta\hat\rho e^{\sigma t}e^{i{\bf q}\cdot {\bf r}}$ and $\delta c\sim \delta\hat c e^{\sigma t}e^{i{\bf q}\cdot {\bf r}}$, we get
\begin{eqnarray}
\label{dis6}
(F-\epsilon\sigma)\delta\hat c+f(c)\delta\hat\rho=0,
\end{eqnarray}
\begin{eqnarray}
\label{dis7}
D_{1}q^{2}\delta\hat c-(D_{2}q^{2}+\sigma(\sigma+\xi))\delta\hat\rho=0,
\end{eqnarray}
where $F=f'(c)\rho-\overline{k}-q^{2}D$. These equations have
non-trivial solutions only if the determinant of the system  is
equal to zero yielding the dispersion relation
\begin{eqnarray}
\label{dis8}
\epsilon\sigma^{3}+(\epsilon\xi-F)\sigma^{2}-(F\xi-\epsilon q^{2}D_{2})\sigma\nonumber\\
-q^{2}(f(c)D_{1}+D_{2}F)=0.
\end{eqnarray}
The condition of marginal stability ($\sigma=0$) corresponds to $f(c)D_{1}+D_{2}F=0$ and the condition of instability is
\begin{eqnarray}
\label{dis9}
f(c)D_{1}+D_{2}F>0.
\end{eqnarray}
We note that the instability criterion does not depend on the value of
$\epsilon$ and $\xi$. Let us consider in detail some particular cases.

\subsection{The case $\epsilon=0$} \label{sec_eps}

When the chemical has a large diffusivity, the temporal term in
Eq. (\ref{intro5}) can be neglected \cite{jager}. Thus, we formally consider
$\epsilon=0$. In that case, the dispersion relation
(\ref{dis8}) reduces to
\begin{eqnarray}
\label{eps1}
\sigma^{2}+\xi\sigma+q^{2}\left (D_{2}+\frac{f(c)D_{1}}{F}\right )=0.
\end{eqnarray}
The discriminant is
\begin{eqnarray}
\label{eps2}
\Delta(q)=\xi^{2}-4 q^{2}\left (D_{2}+\frac{f(c)D_{1}}{F(q)}\right ),
\end{eqnarray}
and the two roots are
\begin{eqnarray}
\label{eps3}
\sigma_{\pm}=\frac{-\xi\pm \sqrt{\Delta(q)}}{2}.
\end{eqnarray}
If $R_{e}(\sigma)<0$ the perturbation decays exponentially rapidly and if
$R_{e}(\sigma)>0$ the perturbation grows exponentially rapidly. In that
case, the system is unstable and $R_{e}(\sigma)$ is the growth rate of the
perturbation. If $D_{2}+f(c)D_{1}/F<0$, then $\Delta> \xi^{2}$ so
that $\sigma_{+}> 0$ (unstable). Alternatively, if
$D_{2}+f(c)D_{1}/F\ge 0$, then $\Delta\le \xi^{2}$. Either
$\Delta\le 0$ and $R_{e}(\sigma)=-\xi/2\le 0$ (stable) or $0\le
\Delta\le \xi^{2}$ and $\sigma_{\pm}\le 0$ (stable). Therefore, the
system is unstable if
\begin{eqnarray}
\label{eps4}
D_{2}<\frac{f(c)D_{1}}{\overline{k}-f'(c)\rho+Dq^{2}},
\end{eqnarray}
and stable otherwise. To determine the range of unstable
wavelengths, we must consider different cases:

\subsubsection{If $\overline{k}-f'(c)\rho\ge 0$:} 
\label{sec_pr}

\begin{figure}
\centering
\includegraphics[width=8cm]{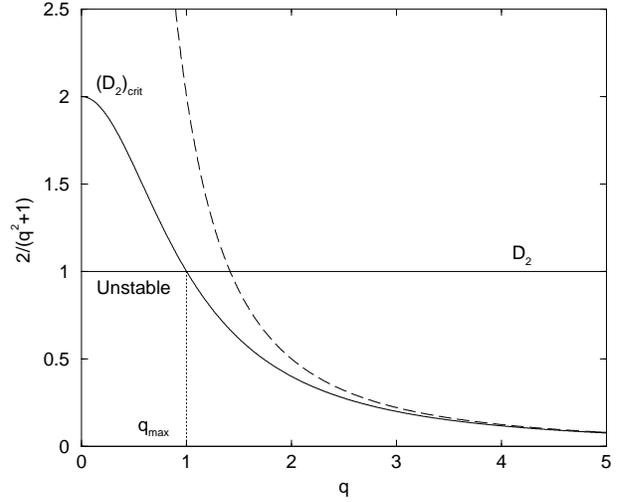}
\caption{Graphical construction determining the range of unstable wavenumbers. We have taken $\xi=D_{2}=D=1$, $fD_{1}=2$ and $\overline{k}-f'(c)\rho=1$ (solid line) or $\overline{k}-f'(c)\rho=0$ (dashed line). }
\label{qcrit}
\end{figure}

\begin{figure}
\centering
\includegraphics[width=8cm]{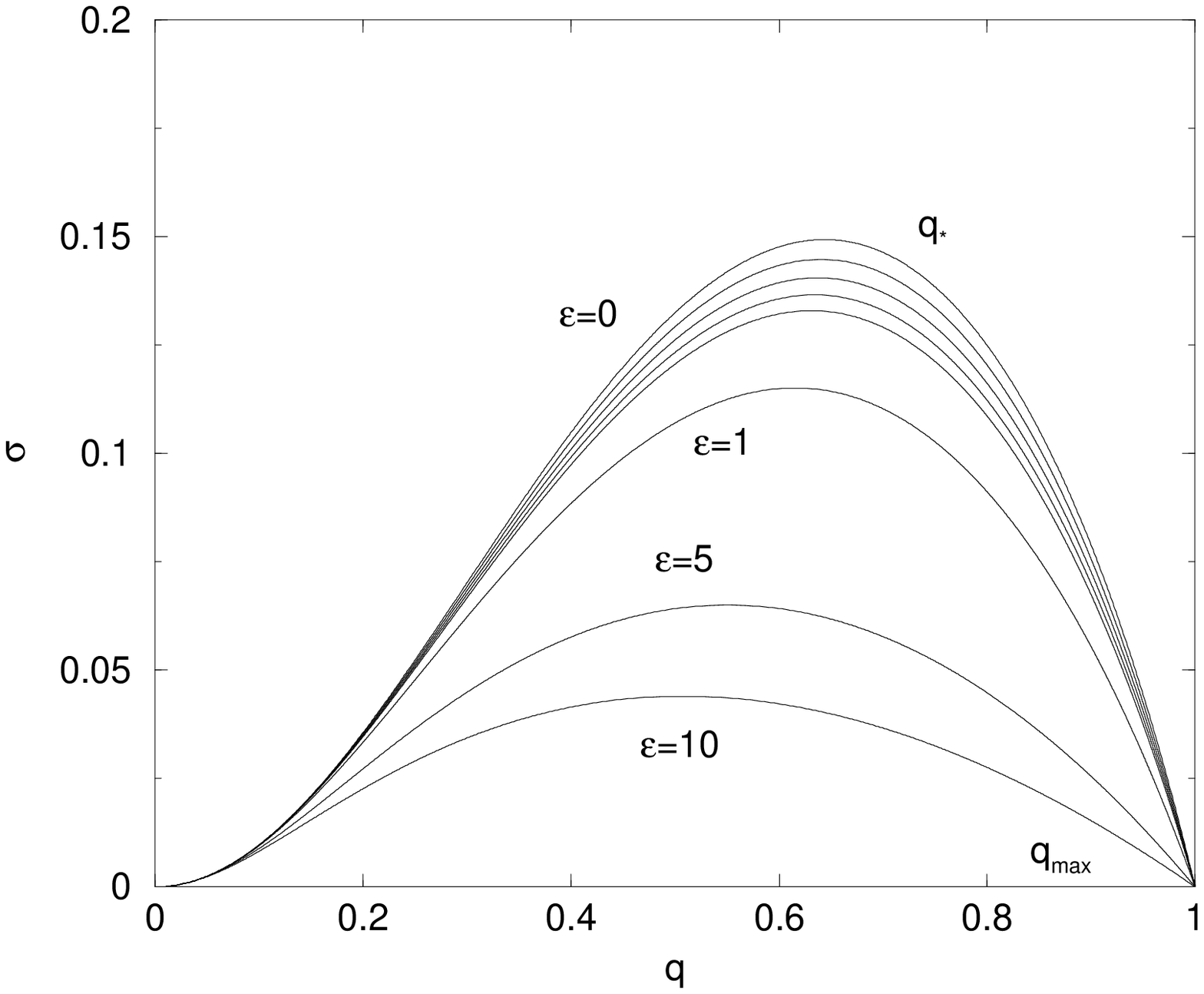}
\caption{Evolution of the growth rate of the perturbation as a function of the wavenumber for different values of $\epsilon$. We have taken $\xi=D_{2}=D=1$, $fD_{1}=2$ and $\overline{k}-f'(c)\rho=1$. }
\label{sigmaq}
\end{figure}

In that case, a {\it necessary} condition of instability is that
\begin{eqnarray}
\label{eps5}
D_{2}<\frac{f(c)D_{1}}{\overline{k}-f'(c)\rho}\equiv (D_{2})_{crit}.
\end{eqnarray}
If this condition is fulfilled, the unstable wavenumbers are
determined by
\begin{eqnarray}
\label{eps6}
q^{2}\le \frac{1}{D}\left\lbrack \frac{f(c)D_{1}}{D_{2}}+f'(c)\rho-\overline{k}\right\rbrack\equiv q_{max}^{2}.
\end{eqnarray}
The growth rate of the perturbation with wavenumber $q$ is
$\sigma_{+}=\frac{1}{2}(-\xi+\sqrt{\Delta(q)})$. Therefore, the most
unstable mode $q_*$ is the one which maximizes $\Delta(q)$. It is
given by
\begin{eqnarray}
\label{eps7}
D q_{*}^{2}=\left\lbrack \frac{f(c)D_{1}(\overline{k}-f'(c)\rho)}{D_{2}}\right\rbrack^{1/2}+f'(c)\rho-\overline{k}.
\end{eqnarray}
The largest growth rate $\sigma_*$  is then determined by
\begin{eqnarray}
\label{eps8}
2\sigma_{*}=-\xi+\sqrt{\xi^{2}+\frac{4f(c)D_{1}}{D}\left (1-\sqrt{\frac{D_{2}(\overline{k}-f'(c)\rho)}{f(c)D_{1}}}\right )^{2}}. \nonumber\\
\end{eqnarray}
In particular, for $\xi=0$, we have
\begin{eqnarray}
\label{eps8b}
2\sigma_{*}=\sqrt{\frac{4f(c)D_{1}}{D}}\left (1-\sqrt{\frac{D_{2}(\overline{k}-f'(c)\rho)}{f(c)D_{1}}}\right ). 
\end{eqnarray}
The range of unstable wavelengths is determined graphically in
Fig. \ref{qcrit} and the evolution of the growth rate of the
perturbation as a function of the wavenumber is plotted in
Fig. \ref{sigmaq} (we have also consider the case $\epsilon\neq 0$ in
this Figure by solving Eq. (\ref{dis8}) which is a second degree
equation in $x=q^{2}$).

\subsubsection{If $\overline{k}-f'(c)\rho=0$:} 

In that case, $(D_{2})_{crit}=+\infty$. The unstable wavenumbers are
determined by
\begin{eqnarray}
\label{eps9}
q^{2}\le \frac{f(c)D_{1}}{D D_{2}}\equiv q_{max}^{2}.
\end{eqnarray}
The most unstable mode is $q_{*}=0$ and the largest growth rate
$\sigma_*$  is given by
\begin{eqnarray}
\label{eps10}
2\sigma_{*}=-\xi+\sqrt{\xi^{2}+\frac{4f(c)D_{1}}{D}}. 
\end{eqnarray}

\subsubsection{If $\overline{k}-f'(c)\rho< 0$:} 
\label{sec_spe}

In that case, the system is unstable for
\begin{eqnarray}
\label{eps11}
\frac{f'(c)\rho-\overline{k}}{D}\le q^{2}\le \frac{1}{D}\left\lbrack \frac{f(c)D_{1}}{D_{2}}+f'(c)\rho-\overline{k}\right\rbrack.
\end{eqnarray}
The growth rate diverges when
\begin{eqnarray}
\label{eps12}
q^{2}\rightarrow \frac{f'(c)\rho-\overline{k}}{D}\equiv q_{0}^{2},
\end{eqnarray}
corresponding to $F(q)=0$. Close to the critical wavenumber $q_{0}$, we have
\begin{eqnarray}
\label{eps13}
\sigma\sim \left (\frac{q_{0}f(c)D_{1}}{2D}\right )^{1/2}\frac{1}{\sqrt{q-q_{0}}}, \quad (q\rightarrow q_{0}^{+}).
\end{eqnarray}
This expression is valid for $\xi$ finite and $\epsilon=0$. It can be
directly obtain from Eq. (\ref{eps1}) by using $F\sim
-2Dq_{0}(q-q_{0})\rightarrow 0$ for $q\rightarrow q_{0}$.  Thus, when
the temporal term is neglected in Eq. (\ref{intro5}),
i.e. $\epsilon=0$, a critical behaviour occurs. This critical
behaviour is regularized for $\epsilon\neq 0$. Indeed, taking
$q=q_{0}$, i.e. $F=0$, in Eq. (\ref{dis8}), we obtain
\begin{eqnarray}
\label{eps14}
\epsilon\sigma^{3}+\epsilon\xi\sigma^{2}+\epsilon q_{0}^{2}D_{2}\sigma-q_{0}^{2}f(c)D_{1}=0.
\end{eqnarray}
Taking the limit $\epsilon\rightarrow 0$, we find that
\begin{eqnarray}
\label{eps15}
\sigma(q_{0})\sim \left (\frac{q_{0}^{2}f(c)D_{1}}{\epsilon}\right )^{1/3},
\end{eqnarray}
which is finite for $\epsilon>0$ but diverges like $\epsilon^{-1/3}$
when $\epsilon\rightarrow 0$.  For $\epsilon\rightarrow 0$ and
$q\rightarrow q_{0}$, the dispersion relation can be simplified in
\begin{eqnarray}
\label{eps16}
\epsilon\sigma^{3}+2Dq_{0}(q-q_{0})\sigma^{2}-q_{0}^{2}f(c)D_{1}=0.
\end{eqnarray}
For $\epsilon=0$ we recover Eq. (\ref{eps13}) and for $q=q_{0}$ we recover Eq. (\ref{eps15}). For $q\rightarrow q_{0}$, we can easily express $q$ as a function of $\sigma$ according to
\begin{eqnarray}
\label{eps16a}
q-q_{0}=\frac{q_{0}^{2}f(c)D_{1}-\epsilon\sigma^{3}}{2D q_{0}\sigma^{2}}.
\end{eqnarray}
On the other hand, for $q=0$, Eq. (\ref{dis8}) reduces to
\begin{eqnarray}
\label{eps16b}
\epsilon\sigma^{3}+(\xi\epsilon-f'(c)\rho+\overline{k})\sigma^{2}-\xi (f'(c)\rho-\overline{k})\sigma =0.
\end{eqnarray}
The positive root of this equation is
\begin{eqnarray}
\label{eps16c}
\sigma(0)=\frac{f'(c)\rho-\overline{k}}{\epsilon}=\frac{Dq_{0}^{2}}{\epsilon},
\end{eqnarray}
which is finite for $\epsilon>0$ but diverges like $\epsilon^{-1}$
when $\epsilon\rightarrow 0$. 

\begin{figure}
\centering
\includegraphics[width=8cm]{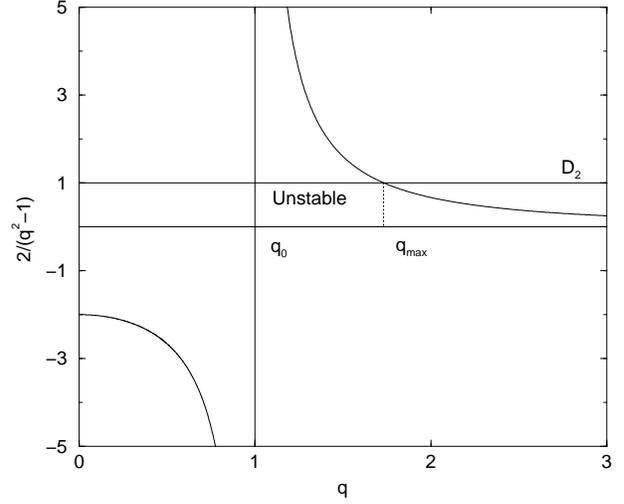}
\caption{Graphical construction determining the range of unstable wavenumbers. We have taken $\xi=D_{2}=D=1$, $fD_{1}=2$ and $\overline{k}-f'(c)\rho=-1$. }
\label{qcritsing}
\end{figure}

\begin{figure}
\centering
\includegraphics[width=8cm]{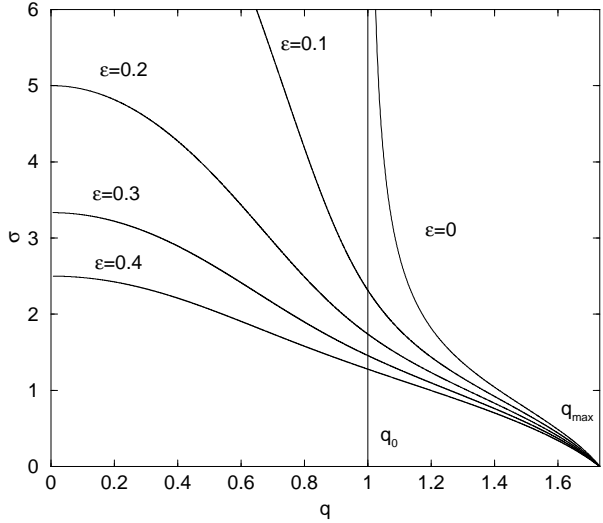}
\caption{Evolution of the growth rate of the perturbation as a function of the wavenumber. We have taken $\xi=D_{2}=D=1$, $fD_{1}=2$ and $\overline{k}-f'(c)\rho=-1$.  }
\label{sigmaqcrit}
\end{figure}

The range of unstable wavelengths is determined graphically in
Fig. \ref{qcritsing} and the evolution of the growth rate of the
perturbation as a function of the wavenumber is plotted in
Fig. \ref{sigmaqcrit} (we have also consider the case $\epsilon\neq 0$
in this Figure by solving Eq. (\ref{dis8})). We see that the case
$\epsilon=0$ is very special. For $\epsilon=0$, the range of
wavenumbers $q<q_{0}$ seems to be stable according to
Eq. (\ref{eps11}) because the two roots of Eq. (\ref{eps1}) are
negative. However, for any finite value of $\epsilon$, a third root
appears. This root is positive and tends to infinity when
$\epsilon\rightarrow 0$ (see Eqs. (\ref{eps15}) and
(\ref{eps16c})). Therefore, this unstable branch
is rejected to infinity when $\epsilon\rightarrow 0$. This implies
that the region $q<q_{0}$ is in fact extremely unstable for
$\epsilon=0^{+}$. In particular, for $\epsilon>0$, the most unstable
mode is $q_{*}=0$ and the largest growth rate is given by
Eq. (\ref{eps16c}).

\subsection{The case $\xi\rightarrow +\infty$} \label{sec_xi}

\begin{figure}
\centering
\includegraphics[width=8cm]{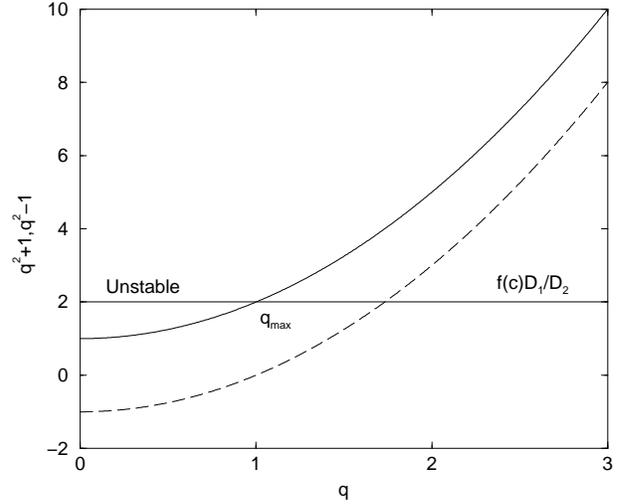}
\caption{Graphical construction determining the range of unstable wavenumbers. We have taken $D_{2}=D=1$, $fD_{1}=2$ and $\overline{k}-f'(c)\rho=1$ (solid line) and $\overline{k}-f'(c)\rho=-1$ (dashed line). }
\label{qcritKS}
\end{figure}

In the overdamped limit, the hydrodynamical equations
(\ref{intro3})-(\ref{intro5}) return the Keller-Segel model
(\ref{intro1})-(\ref{intro2}). Let us consider the stability analysis
in that case for comparison with the inertial case. The dispersion
relation now reads
\begin{eqnarray}
\label{xi1} \epsilon\xi\sigma^{2}-(F\xi-\epsilon q^{2}D_{2})\sigma
-q^{2}(f(c)D_{1}+D_{2}F)=0,
\end{eqnarray}
and the two roots  are
\begin{eqnarray}
\label{xi2}
\sigma_{\pm}=\frac{F\xi-\epsilon D_{2}q^{2}\pm \sqrt{\Delta(q)}}{2\epsilon \xi},
\end{eqnarray}
where
\begin{eqnarray}
\label{xi3}
\Delta(q)=(F\xi+\epsilon q^{2}D_{2})^{2}+4\epsilon\xi q^{2}f(c)D_{1}\ge 0.
\end{eqnarray}
Writing the solution in the form
\begin{eqnarray}
\label{xi4}
\sigma=\frac{-b\pm\sqrt{b^{2}-4 ac}}{2a},
\end{eqnarray}
it is easy to see that the system is stable if (i) $F<\epsilon
q^{2}D_{2}/\xi$ ($b>0$) and if (ii) $F<-f(c)D_{1}/D_{2}$ ($c>0$). It
is unstable otherwise. Since (ii) implies (i), the system is stable
if $F<-f(c)D_{1}/D_{2}$ and unstable otherwise. Thus, the system is
unstable if
\begin{eqnarray}
\label{xi5}
\overline{k}-f'(c)\rho+Dq^{2}<\frac{f(c)D_{1}}{D_{2}},
\end{eqnarray}
and stable otherwise. The range of unstable wavelengths is determined
graphically in Fig. \ref{qcritKS}.

\subsubsection{If $\overline{k}-f'(c)\rho\ge 0$:} 

In that case, the system is unstable for
\begin{eqnarray}
\label{xi6}
D_{2}<\frac{f(c)D_{1}}{\overline{k}-f'(c)\rho+Dq^{2}},
\end{eqnarray}
and stable otherwise. This is the same criterion as for the inertial
model. A {\it necessary} condition of instability is that
\begin{eqnarray}
\label{xi7}
D_{2}<\frac{f(c)D_{1}}{\overline{k}-f'(c)\rho}\equiv (D_{2})_{crit}.
\end{eqnarray}
If this condition is fulfilled, the unstable wavenumbers are such
that
\begin{eqnarray}
\label{xi8}
q^{2}\le \frac{1}{D}\left\lbrack \frac{f(c)D_{1}}{D_{2}}+f'(c)\rho-\overline{k}\right\rbrack\equiv q_{max}^{2}.
\end{eqnarray}
These results are unchanged with respect to the inertial case.

\begin{figure}
\centering
\includegraphics[width=8cm]{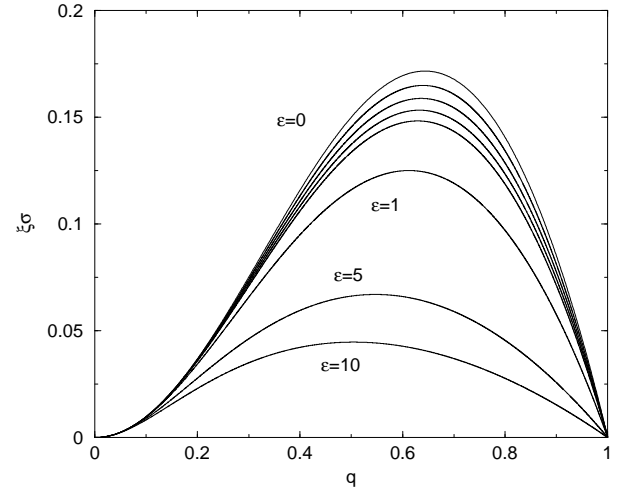}
\caption{Evolution of the growth rate of the perturbation as a function of the wavenumber for the Keller-Segel model. We have taken $D_{2}=D=1$, $fD_{1}=2$ and $\overline{k}-f'(c)\rho=1$.  }
\label{kellerBOSSE}
\end{figure}

We now determine the value of the optimal (most unstable) wavenumber
$q_{*}$ and the corresponding growth rate $\sigma_{*}$ (see
Fig. \ref{kellerBOSSE}). Eq. (\ref{xi1}) is a second order equation in
$\sigma$ whose solutions are given by Eq. (\ref{xi2}). We can maximize
$\sigma_{+}(q)$ to obtain $q_{*}$ and $\sigma_{*}$. However, it appears
simpler to proceed differently. Eq. (\ref{xi1}) can also be viewed as
a second order equation in $x=q^{2}$ of the form
\begin{eqnarray}
\label{xi8a}
A x^{2}+B(\sigma) x+C(\sigma)=0,
\end{eqnarray}
with
\begin{eqnarray}
\label{xi8b}
A=D D_{2},
\end{eqnarray}
\begin{eqnarray}
\label{xi8c}
B(\sigma)=(D\xi+\epsilon D_{2})\sigma-f(c)D_{1}-D_{2}(f'(c)\rho-\overline{k}),
\end{eqnarray}
\begin{eqnarray}
\label{xi8d}
C(\sigma)=\epsilon\xi\sigma^{2}-\xi(f'(c)\rho-\overline{k})\sigma\ge 0.
\end{eqnarray}
There will be two roots  $x_{1}$ and $x_{2}$ provided that $B(\sigma)<0$ and $\Delta(\sigma)=B^{2}-4AC\ge 0$. This last condition can be written
\begin{eqnarray}
\label{xi8e}
\Delta(\sigma)\equiv a\sigma^{2}+b\sigma+c\ge 0,
\end{eqnarray}
with 
\begin{eqnarray}
\label{xi8f}
a=(D\xi-\epsilon D_{2})^{2},
\end{eqnarray}
\begin{eqnarray}
\label{xi8g}
b=-2\lbrack D_{2}(f(c)D_{1}+D_{2}(f'(c)\rho-\overline{k}))\epsilon\nonumber\\
+D\xi f(c)D_{1}-D D_{2}\xi (f'(c)\rho-\overline{k})\rbrack<0,
\end{eqnarray}
\begin{eqnarray}
\label{xi8h}
c=\lbrack f(c)D_{1}+D_{2}(f'(c)\rho-\overline{k})\rbrack^{2}.
\end{eqnarray}
The discriminant $\delta=b^2-4ac$ of Eq. (\ref{xi8e}) is given by 
\begin{eqnarray}
\label{xi8i}
\delta=16 f(c)\xi D D_{1}D_{2}\lbrack (f(c)D_{1}\nonumber\\
+D_{2}(f'(c)\rho-\overline{k}))\epsilon-D\xi (f'(c)\rho-\overline{k})\rbrack.
\end{eqnarray}
The condition $\Delta(\sigma)\ge 0$ to have two roots $x_{1}$ and $x_{2}$ is equivalent to $\sigma\le \sigma_{*}$ with
\begin{eqnarray}
\label{xi8j}
\sigma_{*}=\frac{-b-\sqrt{\delta}}{2a}.
\end{eqnarray}
(Note that the possibility $\sigma\ge (-b+\sqrt{\delta})/2a$
must be rejected since it does not satisfy the requirement
$B(\sigma)<0$).  For $\sigma=\sigma_{*}$, the two roots
$x_{1}=x_{2}=x_{*}$ coincide $(\Delta=0)$ so that $\sigma_{*}$ is the
maximum growth rate. It is reached for an optimal wavenumber
$x_{*}=-B/(2A)$, i.e.
\begin{eqnarray}
\label{xi8k}
q_{*}^{2}=\frac{-B(\sigma_{*})}{2A}.
\end{eqnarray}

\subsubsection{If $\overline{k}-f'(c)\rho <0$:} 

\begin{figure}
\centering
\includegraphics[width=8cm]{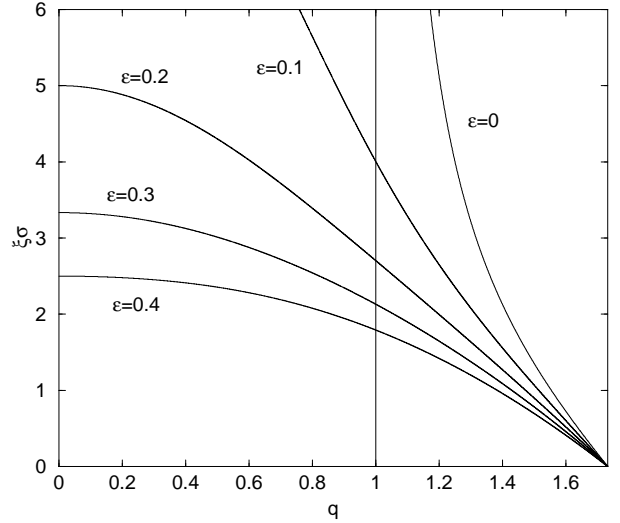}
\caption{Evolution of the growth rate of the perturbation as a function of the wavenumber for the Keller-Segel model. We have taken $D_{2}=D=1$, $fD_{1}=2$ and $\overline{k}-f'(c)\rho=-1$. }
\label{keller}
\end{figure}

In that case the system is unstable for the wavenumbers
\begin{eqnarray}
\label{xi9}
q^{2}\le \frac{1}{D}\left\lbrack \frac{f(c)D_{1}}{D_{2}}+f'(c)\rho-\overline{k}\right\rbrack\equiv q_{max}^{2}.
\end{eqnarray}
The growth rate of the perturbation as a function of the wavenumber is
plotted in Fig. \ref{keller}.  As discussed in Sec. \ref{sec_spe}, the
case $\epsilon=0$ is special and will be considered
specifically in the next section.

\subsection{The case  $\xi\rightarrow +\infty$ and $\epsilon=0$}
\label{sec_mixed}

If we neglect the temporal term $(\epsilon=0)$ in the Keller-Segel model ($\xi\rightarrow +\infty$), we obtain the dispersion relation 
\begin{eqnarray}
\label{mixed1} F\xi\sigma
+q^{2}(f(c)D_{1}+D_{2}F)=0,
\end{eqnarray}
so that $\sigma$ is explicitly given by
\begin{eqnarray}
\label{mixed2} \xi\sigma=q^{2}\left (\frac{f(c)D_{1}}{Dq^{2}+\overline{k}-f'(c)\rho}-D_{2}\right ).
\end{eqnarray}
The instability criterion is given by Eq. (\ref{eps4}).

\subsubsection{If $\overline{k}-f'(c)\rho\ge 0$:} 

This is a particular case of Sec. \ref{sec_pr} corresponding to
$\xi\rightarrow +\infty$. The expression of the largest growth rate is
given by
\begin{eqnarray}
\label{mixed3} \sigma_{*}=\frac{f(c)D_{1}}{\xi D}\left (1-\sqrt{\frac{D_{2}(\overline{k}-f'(c)\rho)}{f(c)D_{1}}}\right )^{2}.
\end{eqnarray}
The other results are unchanged.
 
\subsubsection{If $\overline{k}-f'(c)\rho <0$:} 

The system is unstable for the wavenumbers determined by
Eq. (\ref{eps11}). For $q\rightarrow q_{0}^{+}$, corresponding to $F\sim
-2Dq_{0}(q-q_{0})\rightarrow 0$, the growth rate diverges like
\begin{eqnarray}
\label{mixed4}\xi\sigma\sim \frac{q_{0}f(c)D_{1}}{2D(q-q_{0})}.
\end{eqnarray} 
This divergence is regularized if $\epsilon\neq 0$. Taking $q=q_{0}$, i.e. $F=0$ in Eq. (\ref{xi1}), we get
\begin{eqnarray}
\label{mixed5}\epsilon\xi\sigma^{2}+\epsilon q_{0}^{2}D_{2}\sigma-q_{0}^{2}f(c)D_{1}=0.
\end{eqnarray} 
For $\epsilon\rightarrow 0$, we obtain
\begin{eqnarray}
\label{mixed6}
\sigma(q_{0})\sim \left (\frac{q_{0}^{2}f(c)D_{1}}{\xi\epsilon}\right )^{1/2},
\end{eqnarray} 
which is finite for $\epsilon>0$ but diverges like $\epsilon^{-1/2}$
when $\epsilon\rightarrow 0$.  For $\epsilon\rightarrow
0$ and $q\rightarrow q_{0}^{+}$, Eq. (\ref{xi1}) can be simplified in
\begin{eqnarray}
\label{mixed7}
\epsilon\xi\sigma^{2}+2Dq_{0}\xi (q-q_{0})\sigma-q_{0}^{2}f(c)D_{1}=0.
\end{eqnarray} 
For $\epsilon=0$ we recover Eq. (\ref{mixed4}) and for $q=q_{0}$ we recover Eq. (\ref{mixed6}). The solution of Eq. (\ref{mixed7}) is 
\begin{eqnarray}
\label{mixed8}
\epsilon\sigma=-Dq_{0}(q-q_{0})+\sqrt{D^{2}q_{0}^{2}(q-q_{0})^{2}+\frac{\epsilon q_{0}^{2}f(c)D_{1}}{\xi}}.\nonumber\\
\end{eqnarray} 
On the other hand, for $q=0$, Eq. (\ref{xi1}) leads to
\begin{eqnarray}
\label{mixed9} \sigma(0)=\frac{f'(c)\rho-\overline{k}}{\epsilon},
\end{eqnarray}
which is finite for $\epsilon>0$ but diverges like $\epsilon^{-1}$
when $\epsilon\rightarrow 0$.

Equations (\ref{eps13})-(\ref{eps15}) and
Eqs. (\ref{mixed4})-(\ref{mixed6}) differ because both $\sigma$ and
$\xi$ tend to infinity, so that the expression depend on how the
limits are taken. The general case can be treated as follows. For
$\epsilon=0$, taking $q\rightarrow q_{0}^{+}$ in Eq. (\ref{eps1}) we get
\begin{eqnarray}
\label{mixed10}
\sigma(\sigma+\xi)\sim \frac{q_{0}f(c)D_{1}}{2D(q-q_{0})}.
\end{eqnarray} 
On the other hand, for $\epsilon\neq 0$, taking $q=q_{0}$ (i.e. $F=0$) in Eq. (\ref{dis8}), we get
\begin{eqnarray}
\label{mixed11}
\sigma^{2}(\sigma+\xi)\sim \frac{q_{0}f(c)D_{1}}{\epsilon}.
\end{eqnarray}
Finally, for $\epsilon\rightarrow 0$ and $q\rightarrow q_{0}$, we have
\begin{eqnarray}
\label{mixed12}
\epsilon\sigma^{3}+(\epsilon\xi+2Dq_{0}(q-q_{0}))\sigma^{2}\nonumber\\
+2Dq_{0}(q-q_{0})\xi\sigma-q_{0}^{2}f(c)D_{1}=0.
\end{eqnarray}
which reproduces the correct behaviours (\ref{mixed10}) and (\ref{mixed11}).

\section{Analogy with the Jeans problem in astrophysics}
\label{sec_connect}

\subsection{The damped Euler equations}
\label{sec_damped}

Let us consider a particular case of Eqs.
(\ref{intro3})-(\ref{intro5}) corresponding to $D_2(\rho,c)=p'(\rho)$
and $D_1(\rho,c)=\rho S'(c)$ where $p$ and $S$ are arbitrary
functions. In that case, the hydrodynamical
equations take the form
\begin{eqnarray}
\label{damped1}
\frac{\partial\rho}{\partial t}+\nabla\cdot (\rho {\bf u})=0,
\end{eqnarray}
\begin{eqnarray}
\label{damped2}
\frac{\partial {\bf u}}{\partial t}+({\bf u}\cdot \nabla){\bf u}=-\frac{1}{\rho}\nabla p+\nabla S(c)-\xi {\bf u},
\end{eqnarray}
\begin{eqnarray}
\label{damped3}
\epsilon\frac{\partial c}{\partial t}=-k(c)c+\rho f(c)+D\Delta c.
\end{eqnarray}
For $\xi\rightarrow +\infty$, we can neglect the inertia of the
particles so that $\rho {\bf u}\simeq -\frac{1}{\xi}(\nabla
p-\rho\nabla S(c))$. Substituting this relation in
Eq. (\ref{damped1}), we obtain a special case of the Keller-Segel
model
\begin{eqnarray}
\label{damped4}
\frac{\partial\rho}{\partial t}=\nabla\cdot \left \lbrack \chi\left( \nabla p-\rho S'(c)\nabla c\right )\right \rbrack,
\end{eqnarray}
where we have set $\chi=1/\xi$. Equations
(\ref{damped1})-(\ref{damped2}) can be viewed as fluid equations
appropriate to the chemotactic problem. Equation (\ref{damped1}) is an
equation of continuity and Eq. (\ref{damped2}) is similar to the Euler
equation where $p$ plays the role of a pressure and the chemotactic
attraction plays the role of a force. Since $p=p(\rho)$, these equations
describe a barotropic fluid. The main novelty of these equations with
respect to usual hydrodynamical equations is the presence of a
friction force which allows to make a connection between hyperbolic
($\xi=0$) and parabolic ($\xi\rightarrow +\infty$) models.

This hydrodynamic model including a friction force is similar to the
damped barotropic Euler-Poisson system which describes a gas of
self-gravitating Brownian particles \cite{gt,virial} or the violent
relaxation of collisionless stellar systems on the coarse-grained
scale in astrophysics \cite{csr}. In that analogy, the concentration
of the chemical $c$ plays the role of the gravitational potential
$\Phi$.  The main difference between the two models is that the
Poisson equation for self-gravitating systems is replaced by a more
general field equation (\ref{damped3}) for bacterial populations. To
emphasize the connection with astrophysical problems, let us consider
a particular case of Eqs. (\ref{damped1})-(\ref{damped3}) where
$\epsilon=0$, $S(c)=c$ and $k$ and $f$ are constant. Then, introducing
notations similar to those used in astrophysics (noting $c=-\Phi$,
$k/D=k_0^2$, $f/D=S_d G$), we can rewrite
Eqs. (\ref{damped1})-(\ref{damped3}) in the form
\begin{eqnarray}
\label{damped5}
\frac{\partial\rho}{\partial t}+\nabla\cdot (\rho {\bf u})=0,
\end{eqnarray}
\begin{eqnarray}
\label{damped6}
\frac{\partial {\bf u}}{\partial t}+({\bf u}\cdot \nabla){\bf u}=-\frac{1}{\rho}\nabla p-\nabla \Phi-\xi {\bf u},
\end{eqnarray}
\begin{eqnarray}
\label{damped7}
\Delta\Phi-k_{0}^{2}\Phi=S_{d} G(\rho-\overline{\rho}).
\end{eqnarray}
When $k_0=\overline{\rho}=0$, these equations are isomorphic to the
damped Euler-Poisson system describing self-gravitating Brownian
particles \cite{gt,virial}. In the strong friction limit, we get
\begin{eqnarray}
\label{damped8} \frac{\partial\rho}{\partial t}=\nabla\cdot \left
\lbrack \frac{1}{\xi}\left( \nabla p+\rho\nabla \Phi\right )\right
\rbrack,
\end{eqnarray}
which can be interpreted as a generalized Smoluchowski equation.
Thus, for $\xi\rightarrow +\infty$, we obtain the generalized
Smoluchowski-Poisson system describing self-gravitating Brownian
particles in an overdamped limit \cite{gt}. Alternatively, for $\xi=0$ we
recover the barotropic Euler-Poisson system that has been studied at
length in astrophysics to determine the period of stellar pulsations
\cite{cox} and the formation of large-scale structures in cosmology
\cite{peebles}. Therefore, the chemotactic model
(\ref{damped5})-(\ref{damped7}) is similar to astrophysical models
with additional terms. In the astrophysical context, the case
$k_{0}\neq 0$ would correspond to a shielding of the gravitational
interaction on a typical length $k_{0}^{-1}$. This Yukawa shielding
appears in theories where the graviton has a mass but in that case
$k_0$ is very small which does not need to be the case in the
biological problem.

We now consider the linear dynamical stability of an infinite and
homogeneous solution of Eqs.  (\ref{damped5})-(\ref{damped7}). By
mapping the equations (\ref{damped5})-(\ref{damped7}) onto a
generalized astrophysical model, we shall see that the instability
criteria obtained in Sec. \ref{sec_instinert} are connected to (and
extend) the Jeans instability criterion of astrophysics
\cite{jeans}. To avoid the Jeans swindle \cite{bt} when $k_{0}=0$, we have
introduced a ``neutralizing background'' $\overline{\rho}$ in
Eq. (\ref{damped7}). In fact, in the biological problem, this term
appears {\it naturally} when we consider the limit of large
diffusivity of the chemical (see \cite{jager}); therefore, there is no
``Jeans swindle'' in the biological problem based on
Eq. (\ref{intro2}).  A similar term $\overline{\rho}$ appears in
cosmology when we take into account the expansion of the universe and
work in the comoving frame \cite{peebles}.

\subsection{The dispersion relation}
\label{sec_disj}

We consider an infinite and homogeneous stationary solution of Eqs.
(\ref{damped5})-(\ref{damped7}) with ${\bf u}={\bf 0}$, $\rho={\rm
Cst.}$ and $\Phi={\rm Cst.}$ such that
\begin{eqnarray}
\label{disj1}
-k_{0}^{2}\Phi=S_{d}G(\rho-\overline{\rho}).
\end{eqnarray}
For a pure Newtonian interaction with $k_{0}=0$, we have
$\rho=\overline{\rho}$. Linearizing the equations around this
stationary solution, we get
\begin{eqnarray}
\label{disj2}
\frac{\partial \delta\rho}{\partial t}+\rho \nabla\cdot \delta {\bf u}=0,
\end{eqnarray}
\begin{eqnarray}
\label{disj3}
\rho \frac{\partial \delta{\bf u}}{\partial t}=-c_{s}^{2}\nabla\delta\rho-\rho \nabla \delta \Phi-\xi\rho \delta{\bf u},
\end{eqnarray}
\begin{eqnarray}
\label{disj4}
\Delta\delta\Phi-k_{0}^{2}\delta\Phi=S_{d}G\delta\rho,
\end{eqnarray}
where we have introduced the velocity of sound $c_{s}^{2}=p'(\rho)$. Eliminating the velocity between Eqs. (\ref{disj2}) and (\ref{disj3}), we obtain
\begin{eqnarray}
\label{disj5}
\frac{\partial^{2}\delta\rho}{\partial t^{2}}+\xi\frac{\partial\delta\rho}{\partial t}=c_{s}^{2}\Delta\delta\rho+\rho\Delta\delta \Phi.
\end{eqnarray}
Looking for solutions of the form $\delta\rho\sim \delta\hat\rho e^{i({\bf k}\cdot {\bf r}-\omega t)}$, we get
\begin{eqnarray}
\label{disj6}
(-\omega^{2}-i\xi\omega+c_{s}^{2}k^{2})\delta\hat\rho=-\rho k^{2}\delta\hat\Phi,
\end{eqnarray}
\begin{eqnarray}
\label{disj7}
\delta\hat\Phi=-\frac{S_{d}G}{k^{2}+k_{0}^{2}-i\omega}\delta\hat\rho.
\end{eqnarray}
From these equations, we obtain the dispersion relation
\begin{eqnarray}
\label{disj8}
\omega (\omega+i\xi)=c_{s}^{2}k^{2}-\frac{S_{d}G\rho k^{2}}{k^{2}+k_{0}^{2}}.
\end{eqnarray}
In the case $\xi=0$ and $k_{0}=0$, we recover the usual Jeans
dispersion relation \cite{jeans}:
\begin{eqnarray}
\label{disj9}
\omega^{2}=c_{s}^{2}k^{2}-S_{d}G\rho.
\end{eqnarray}

\subsection{Instability criterion}
\label{sec_icj}

If we set $\sigma=-i\omega$, the dispersion relation becomes
\begin{eqnarray}
\label{icj1}
\sigma^{2}+\xi\sigma+k^{2}\left (c_{s}^{2}-\frac{S_{d}G\rho}{k^{2}+k_{0}^{2}}\right )=0.
\end{eqnarray}
The two roots are
\begin{eqnarray}
\label{icj2}
\sigma=\frac{-\xi\pm \sqrt{\Delta(k)}}{2},
\end{eqnarray}
with
\begin{eqnarray}
\label{icj3}
\Delta(k)=\xi^{2}-4 k^{2}\left (c_{s}^{2}-\frac{S_{d}G\rho}{k^{2}+k_{0}^{2}}\right ).
\end{eqnarray}
Accordingly, the system is unstable if
\begin{eqnarray}
\label{icj4}
c_{s}^{2}<\frac{S_{d}G\rho}{k^{2}+k_{0}^{2}},
\end{eqnarray}
and stable otherwise. A necessary condition of instability is
\begin{eqnarray}
\label{icj5}
c_{s}^{2}<\frac{S_{d}G\rho}{k_{0}^{2}}\equiv (c_{s}^{2})_{crit}.
\end{eqnarray}
If this condition is fulfilled the unstable wavelengths are such that
\begin{eqnarray}
\label{icj6}
k\le \sqrt{\frac{S_{d}G\rho}{c_{s}^{2}}-k_{0}^{2}}\equiv k_{max}.
\end{eqnarray}
The wavelength which has the largest growth rate is given by
\begin{eqnarray}
\label{icj7} k_{*}^{2}=\left (\frac{S_{d}G\rho
k_{0}^{2}}{c_{s}^{2}}\right )^{1/2}-k_{0}^{2},
\end{eqnarray}
and the corresponding growth rate is given by
\begin{eqnarray}
\label{icj8}
2\sigma_{*}=-\xi+\sqrt{\xi^{2}+4S_{d}G\rho \left (1-\sqrt{\frac{c_{s}^{2}k_{0}^{2}}{S_{d}G\rho}}\right )^{2}}. \nonumber\\
\end{eqnarray}

The instability criterion (\ref{icj4}) is equivalent to the
instability criterion (\ref{eps4}) of Sec. \ref{sec_eps} for the
particular case of the chemotactic model considered in
Sec. \ref{sec_damped}. The parallel with astrophysics is interesting
to develop in order to give a more physical interpretation to Eq. 
(\ref{eps4}). All the other formulae can be interpreted accordingly. In
particular, we note that the coefficient $D_2$ plays the role of a
velocity of sound $c_s^2$.

\subsection{Particular cases}
\label{sec_pc}

Let us consider particular cases of the foregoing
expressions:

\noindent $\bullet$ For $c_{s}=0$, one has $k_{max}=+\infty$,
$k_{*}=+\infty$ and
\begin{eqnarray}
\label{pc1}
\sigma_{*}=\frac{-\xi+\sqrt{\xi^{2}+4S_{d}G\rho}}{2}. \nonumber\\
\end{eqnarray}

\noindent $\bullet$ For $k_{0}=0$ (Newtonian potential), one has
$(c_{s}^{2})_{crit}=+\infty$, $k_{max}=\left
({S_{d}G\rho}/{c_{s}^{2}}\right )^{1/2}\equiv k_{J}$ (Jeans length),
$k_{*}=0$ and
\begin{eqnarray}
\label{pc2}
\sigma_{*}=\frac{-\xi+\sqrt{\xi^{2}+4S_{d}G\rho}}{2}. \nonumber\\
\end{eqnarray}

\noindent $\bullet$ For $\xi=0$, one has
\begin{eqnarray}
\label{pc3}
\sigma_{*}=(S_{d}G\rho)^{1/2}\left (1-\sqrt{\frac{c_{s}^{2}k_{0}^{2}}{S_{d}G\rho}}\right ).
\end{eqnarray}

\noindent $\bullet$ For $\xi\rightarrow +\infty$, one has
\begin{eqnarray}
\label{pc4}
\sigma_{*}=\frac{S_{d}G\rho}{\xi} \left (1-\sqrt{\frac{c_{s}^{2}k_{0}^{2}}{S_{d}G\rho}}\right )^{2}.
\end{eqnarray}

\subsection{Isothermal gas}
\label{sec_isog}

\begin{figure}
\centering
\includegraphics[width=8cm]{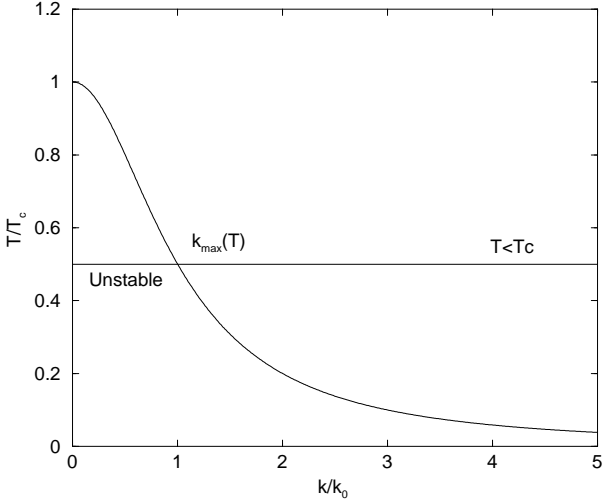}
\caption{Graphical construction determining the range of unstable 
wavenumbers.}
\label{kT}
\end{figure}

\begin{figure}
\centering
\includegraphics[width=8cm]{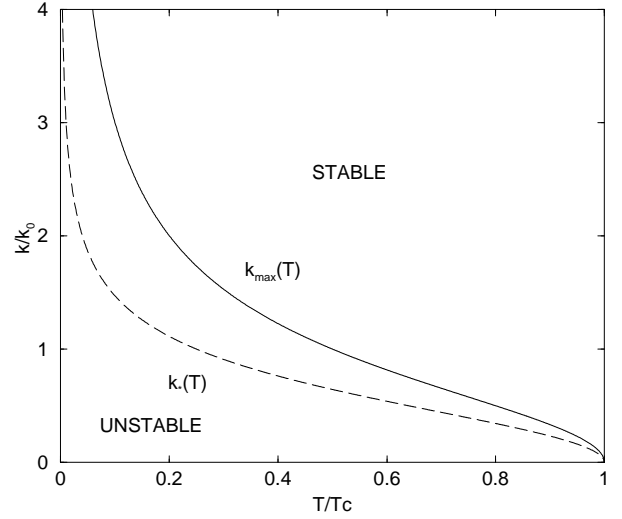}
\caption{Maximum wavenumber $k_{max}(T)$ and most unstable wavenumber $k_{*}(T)$ as a function of the temperature $T$. The line  $k_{max}(T)$ determines the separation between stable and unstable states. }
\label{kmaxT}
\end{figure}

\begin{figure}
\centering
\includegraphics[width=8cm]{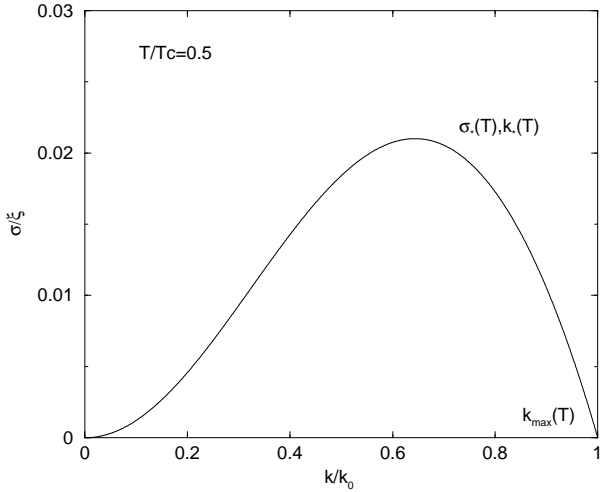}
\caption{Growth rate of the perturbation as a function of the wavenumber. We have taken  $4S_{d}G\rho/\xi^{2}=1$ and $T/T_{c}=0.5$.}
\label{ksigmaT}
\end{figure}

\begin{figure}
\centering
\includegraphics[width=8cm]{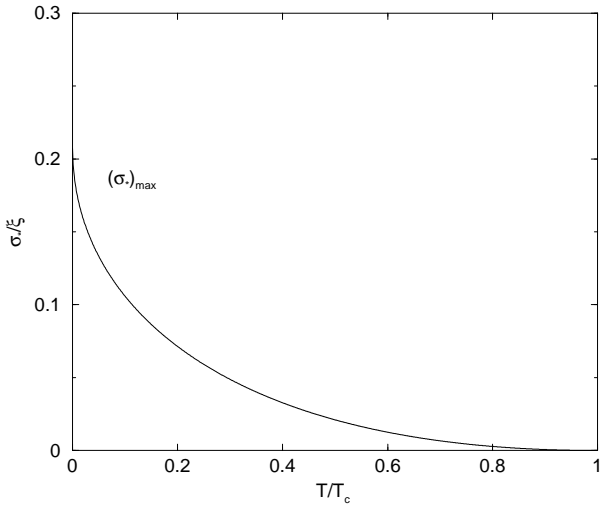}
\caption{Dependence of the largest growth rate $\sigma_{*}(T)$ with the temperature. We have taken   $4S_{d}G\rho/\xi^{2}=1$.}
\label{starT}
\end{figure}

For an isothermal gas with an equation of state $p=\rho T$ (for
simplicity, we have noted $T$ instead of $k_{B}T/m$), the velocity of
sound is equal to the square root of the temperature: $c_{s}^{2}=T$.
It is relevant to re-express the previous relations as
follows. Introducing the critical temperature
\begin{eqnarray}
\label{isog1}
T_{c}=\frac{S_{d}G\rho}{k_{0}^{2}}, 
\end{eqnarray} 
the growth rate of the perturbation can be written
\begin{eqnarray}
\label{isog2}
\frac{2\sigma}{\xi}=-1+\sqrt{1-\frac{4S_{d}G\rho}{\xi^{2}}\frac{k^{2}}{k_{0}^{2}}\left (\frac{T}{T_{c}}-\frac{1}{1+(k/k_{0})^{2}}\right )}.\nonumber\\
\end{eqnarray} 
The condition of instability reads 
\begin{eqnarray}
\label{isog3}
\frac{T}{T_{c}}\le \frac{1}{1+(k/k_{0})^{2}},
\end{eqnarray} 
and a necessary condition of instability is $T<T_{c}$.  For $T<T_{c}$
the unstable wavenumbers (see Fig. \ref{kT}) are such that $k\le
k_{max}(T)$ with
\begin{eqnarray}
\label{isog4}
\frac{k_{max}(T)}{k_{0}}=\sqrt{\frac{T_{c}}{T}-1}.
\end{eqnarray}
The wavenumber with the largest growth rate is given by
\begin{eqnarray}
\label{isog5}
\frac{k_{*}(T)}{k_{0}}=\left\lbrack \left (\frac{T_{c}}{T}\right )^{1/2}-1\right\rbrack^{1/2},
\end{eqnarray}
and the largest growth rate by
\begin{eqnarray}
\label{isog6}
\frac{2\sigma_{*}}{\xi}=-1+\sqrt{1+\frac{4S_{d}G\rho}{\xi^{2}}\left\lbrack 1-\left (\frac{T}{T_{c}}\right )^{1/2}\right\rbrack^{2}}.
\end{eqnarray}
The maximum wavenumber $k_{max}(T)$ and the most unstable wavenumber
$k_{*}(T)$ are plotted as a function of the temperature in
Fig. \ref{kmaxT}. In Fig. \ref{ksigmaT}, we represent the growth rate
$\sigma(k)$ as a function of the wavenumber. Finally, in
Fig. \ref{starT}, we plot the largest growth rate $\sigma_{*}(T)$ as a function of the temperature. The maximum value of the largest growth rate $\sigma_{*}(T)$ is obtained for $T=0$ and is given by 
\begin{eqnarray}
\label{isog7}
\frac{2(\sigma_{*})_{max}}{\xi}=-1+\sqrt{1+\frac{4S_{d}G\rho}{\xi^{2}}}.
\end{eqnarray}

\begin{figure}
\centering
\includegraphics[width=8cm]{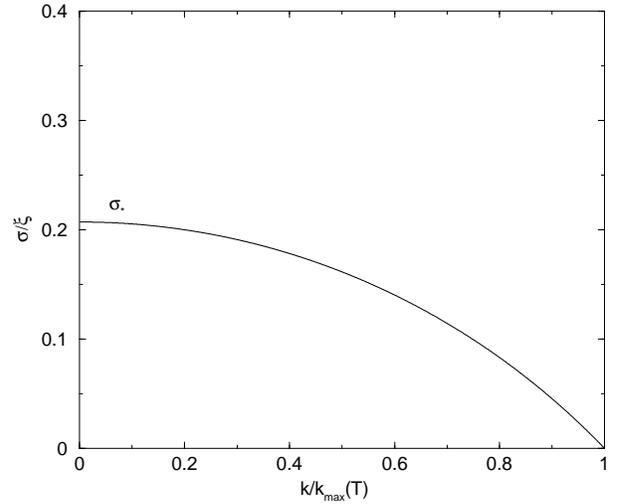}
\caption{Growth rate $\sigma(k)$ as a function of the wavenumber when $k_{0}=0$. We have taken $4S_{d}G\rho/\xi^{2}=1$. In terms of the variable $k/k_{max}(T)$, the curve is independent on the temperature.}
\label{kzero}
\end{figure}

For $k_{0}=0$ (Newtonian interaction), then $T_{c}=+\infty$ and the growth rate of the perturbation can be expressed as
\begin{eqnarray}
\label{isog8}
\frac{2\sigma}{\xi}=-1+\sqrt{1-\frac{4k^{2}}{\xi^{2}}\left ({T}-\frac{S_{d}G\rho}{k^{2}}\right )}.
\end{eqnarray} 
The condition of instablity is 
\begin{eqnarray}
\label{isog9}
k\le k_{max}=\left (\frac{S_{d}G\rho}{T}\right )^{1/2},
\end{eqnarray}
where $k_{max}$ is the equivalent of the Jeans wavenumber.
The maximum growth rate is obtained for $k_{*}=0$ and its value is given by
\begin{eqnarray}
\label{isog10}
\frac{2\sigma_{*}}{\xi}=-1+\sqrt{1+\frac{4S_{d}G\rho}{\xi^{2}}},
\end{eqnarray} 
independently of the temperature. The growth rate $\sigma(k)$ is
represented as a function of the wavenumber in Fig. \ref{kzero}.

\section{Conclusion} \label{sec_conc}

In this paper, we have studied the linear dynamical stability of an
infinite and homogeneous distribution of biological cells whose
density distribution evolves under the process of chemotaxis. We have
modeled their evolution by hydrodynamical equations including a
friction force \cite{gt,crrs}. This inertial model takes into account
the fact that the cells do not respond immediately to the chemotactic
drift but that there is a relaxation time $\xi^{-1}$ for their
velocity to get aligned with the chemotactic gradient. The usual
Keller-Segel model \cite{ks} is recovered in the strong friction limit
$\xi\rightarrow +\infty$ (or for large times $t\gg
\xi^{-1}$). Alternatively, for $\xi=0$, we recover 
the inertial model of
\cite{gamba}. We have shown that these equations were similar to those
describing self-gravitating Brownian particles and that the dynamical
stability of biological populations was related to the Jeans problem
in astrophysics. These results extend the analogies between biology
and astrophysics investigated in \cite{crrs}.

The mathematical model (\ref{intro3})-(\ref{intro5}) considered in
this paper can have applications in biology. Depending on the
value of the parameters, it can decribe different sorts of
systems. The Keller-Segel model (\ref{intro1})-(\ref{intro2}) obtained
in the overdamped limit $\xi\rightarrow \infty$ in which inertial
terms can be neglected is appropriate to describe experiments on
bacteria like {\it Escherichia Coli} and slime mold amoebae like {\it
Dictyostelium discoideum} \cite{ks}. These systems exhibit pointwise
concentrations as a result of chemotactic collapse. On the other hand, the
hydrodynamic model (\ref{ga1})-(\ref{ga3}) was shown to generate a
vascular network starting from randomly seeded endothelial cells
\cite{gamba,filbet}. This can account for experiments of {\it in vitro}
formation of blood vessels where cells randomly spread on a gel matrix
autonomously organize to form a connected network that
can be interpreted as the initiation of angiogenesis. This is also
similar to the formation of capillary blood vessels in
living beings during embriogenesis \cite{embrio}. The authors
of \cite{gamba} evidence a percolative transition as a function of the
concentration of cells. Above a critical density, the system forms a
continuous multi-cellular network which can be described by a
collection of nodes connected by chords. For even higher
concentrations a ``swiss cheese'' pattern is observed. Such structures
can be obtained only if inertial terms are accounted for. Another
hydrodynamic model of bacterial colonies taking into account interial
terms has been proposed by Lega \& Passot
\cite{lega} to describe the evolution of bacterial colonies growing on 
soft agar plates. This model consists in advection-reaction-diffusion
equations for the concentrations of nutrients, water, and bacteria,
coupled to a single hydrodynamic equation for the velocity field of
the bacteria-water mixture. This model is able to reproduce the usual
colony shapes together with nontrivial dynamics inside the colony such
as vortices and jets recently observed in wet colonies of {\it
Bacillus subtilis} \cite{mendelson}. This can be linked to a process
of inverse cascade of energy as in two-dimensional hydrodynamic
turbulence. Lega \& Passot \cite{lega} show that the large-scale
Reynolds numbers can be relatively high so that inertial effects have
to be taken into account to adequately model the experiments of
\cite{mendelson}. It is shown also that viscosity is important in this
model. Although the hydrodynamic equations of \cite{lega} are
different from Eqs. (\ref{intro3})-(\ref{intro5}), their model
displays a mechanism for collective motion towards fresh nutrients
which is similar to classical chemotaxis. In particular, a
chemotacticlike behaviour and a connection to the Keller-Segel model
(\ref{intro1})-(\ref{intro2}) is obtained for short times.

In this paper, we have considered solutions of
Eqs. (\ref{intro3})-(\ref{intro5}) near an infinite and homogeneous
distribution and we have investigated the time dependence
of these solutions in the linear regime \footnote{The linear
dynamical stability of {\it inhomogeneous} distributions of bacteria
has also been studied in
\cite{crs,sc} for overdamped models and in \cite{virial} for inertial
models, when the equation for the concentration of the chemical takes
the form of a Poisson equation (\ref{conc2}) like in gravity. In these
studies, the distribution of particles is self-confined \cite{virial}
or confined in a finite domain (box) \cite{crs,sc}. In biology, the
box can represent a droplet or the container itself.}. When the criterion
(\ref{dis9}) is fulfilled, the appearance of a spontaneous
perturbation can lead to an instability. The perturbation grows until
the system can no longer be described by equilibrium or
near-equilibrium equations. In that case, we must account for the full
nonlinearities encapsulated in Eqs. (\ref{intro3})-(\ref{intro5}). Of
course, the nonlinear regime of instability is the most relevant for
biological applications. This nonlinear regime   has been
investigated in detail for a reduced version of the Keller-Segel model
\cite{jager}:
\begin{eqnarray}
\label{conc1}
\frac{\partial\rho}{\partial t}=D\Delta\rho-\chi\nabla\cdot (\rho\nabla c),
\end{eqnarray}
\begin{eqnarray}
\label{conc2}
\Delta c=-\lambda\rho.
\end{eqnarray}
In that case, the concentration of the chemical is related to the
concentration of the bacteria by a Poisson equation. These equations
are isomorphic to the Smoluchowski-Poisson system
\begin{eqnarray}
\label{conc3}
\frac{\partial\rho}{\partial t}=\nabla \cdot \left\lbrack \frac{1}{\xi}\left (T\nabla\rho+\rho\nabla\Phi\right )\right\rbrack,
\end{eqnarray}
\begin{eqnarray}
\label{conc4}
\Delta \Phi=S_{d}G\rho.
\end{eqnarray} 
describing self-gravitating Brownian particles \cite{crs}. In
dimension $d\ge 2$, they exhibit blow-up solutions leading ultimately
to the formation of Dirac peaks. This corresponds to a chemotactic
collapse in biology or to an isothermal collapse (in the canonical
ensemble) in gravity. There is a vast literature on the theoretical
study of these equations both in applied mathematics (see the review
by Horstmann
\cite{horstmann}) and in physics
\cite{crs,sc,post,time,virial}. Generalized chemotactic models and generalized
gravitational models have also been studied, like in \cite{anomalous}
to account for anomalous diffusion or like in \cite{virial} to account for
inertial effects. On the other hand, bifurcations between ``stripes''
and ``spots'' have been found when the degradation of the secreted
chemical is taken into account so that the equilibrium structures of
the bacterial colonies are similar to ``domain walls'' in phase
ordering kinetics
\cite{wall}. The linear instability regime that we have considered in
this paper initiates the nonlinear regime where interesting and
non-trivial structures form, accounting for the morphogenesis of
bacterial populations. In the linear instability analysis, the general
form of perturbation is a superposition of sinusoidal waves. Each
single wave corresponds to a ``streak'' with relatively high
density. However, other patterns like regularly spaced ``clouds'' can
be obtained by a proper superposition of ``streaks'' (see Appendix A
of \cite{ks}). These ``clouds'' will be presumably selected by
nonlinear effects and each of them can initiate a local collapse
leading to pointwise blow-up \cite{horstmann,crs,sc}. Indeed, these
clouds have the radial symmetry that is assumed at the start in most
studies of chemotactic collapse. This will lead to a set of $N$
singular structures. These compact structures interact with each other
and lead to a coarsening process where the number of clusters decays
in time as they collapse to each other. This process may share some
analogies with the aggregation of vortices in two-dimensional decaying
turbulence \cite{turb}. Therefore, the connection between the linear
regime investigated in this paper and the nonlinear regime
investigated in \cite{horstmann,crs,sc} is relatively clear.

\end{document}